# Isolated zones and granulation of complex network systems

Olexandr Polishchuk

Laboratory of Modeling and Optimization of Complex Systems
Pidstryhach Institute for Applied Problems of Mechanics and Mathematics, National Academy of Sciences of Ukraine,
Lviv, Ukraine
od_polishchuk@ukr.net

**Abstract** – *The problem of blocking certain areas of network systems and their boundary case, which is the granulation of network to a set of isolated zones, is investigated in the article. The losses that await the system due to blocking of its separate components are determined, and method for finding alternative ways of flow motion bypassing the isolated zone of network are proposed. The granulation process of the network system is reflected as a temporal network and the losses that await the system during the deployment of such processes are calculated.*

## INTRODUCTION

In real network systems (NS), processes that need to be stopped quickly are often unfolding: the spread of epidemics, forest fires, agricultural pests, computer viruses, invasion processes [1-5], etc. To prevent the spread of such phenomena by the system, the theory of complex networks usually uses methods of percolation theory [6], which consist in sequential removal of a certain part of network nodes and connections until the percolation cluster breaks up into unconnected components, one of which contains a threat. However, there is another way to solve such problems, which is to purposefully isolate certain parts (zones) of the network. Isolated zones often arise as a result of natural disasters, regional military conflicts, radioactive or chemical contamination of territories [7-11], etc. They appear as a result of damage of complex technological and biological systems [12-14]. From a functional point of view, the isolation of certain subnet of the source complex network (CN) means the complete cessation or significant restriction of flows motion from (to, through) it. This raises a number of issues related to the functioning of NS in general. First, the isolated zone may contain the nodes- generators of flows that need to be replaced. Thus, due to the earthquake and tsunami in Japan on March 11, 2011 in the Fukushima region, almost 40 enterprises - manufacturers of components for the automotive industry were destroyed. As a result, almost all Japanese automakers temporarily stopped their conveyors [15]. Due to the occupation of Donbass, there were problems related to the supply of Group A coal to Ukrainian thermal power plants, etc. Second, the isolated zone may contain the nodes-receivers of flows, which also need to be replaced. Economic sanctions, which are occasionally applied to separate countries, require the search for new markets for products previously supplied to them. In this case, the countries producing these products expect multibillion-dollar losses. Finally, there is the problem of finding alternative routes for transit flows that passed through an isolated zone (difficulties with the transit of Ukrainian products to Central Asia and China and vice versa). That is, blocking a separate subnet of the source network often leads to destabilization of the system operation at a whole [16].

## 1. ISOLATED ZONES AND ALTERNATIVE PATHS OF FLOW MOTION

Assume that the isolated zone of network forms a certain subsystem $\tilde{S}$ of the network system $S$. Determine the integral flow adjacency matrix of the system $S$ by the ratio [17]

$$\mathbf{V}(t) = \{V_{ij}(t)\}_{i,j=1}^{N}, \qquad V_{ij}(t) = \int_{t-T}^{t} v_{ij}(\tau)d\tau, \quad t \geq T > 0.$$

The values of elements of the matrix $\mathbf{V}(t)$ are calculated on the basis of empirical data on the movement of flows in the network, which are easily obtained using modern technologies. Since in each real system usually the flows of different types are moving (passengers and cargo in the transport network), we assume that the elements $V_{ij}(t)$ of matrix $\mathbf{V}(t)$ reflect the financial equivalent of the cost of these flows, i.e. elements $v_{ij}(t)$ are equal to the cost of flow that is on the edge $(n_i, n_j)$, $i,j = \overline{1,N}$, at a moment of time $t \geq T > 0$.

To quantify the real losses or potential risks that await system $S$ in the case of isolation of subsystem $\tilde{S}$, we use the following parameters.

1. The total values of parameters of output influence of the nodes belonging to isolated zone, operate outside it and allow us to evaluate the losses from the need to replace the flows generators that are in this zone. Let us that $H_{\tilde{S}} = \{n_i\}_{i=1}^{\tilde{N}}$ is the set of nodes, belonging to the structure of subsystem $\tilde{S}$. Denote by $G_{\tilde{S}}^{out}$ the set of all nodes-generators of flows belonging to the set $H_{\tilde{S}}$ and determine by means of parameter

$$\xi_{\tilde{S}}^{out}(t) = \sum_{i \in G_{\tilde{S}}^{out}} \xi_i^{out}(t) / s(\mathbf{V}(t))$$

the output influence strength of subsystem $\tilde{S}$ on NS at a whole. Here

$$s(\mathbf{V}(t)) = \sum_{i=1}^{N} \sum_{j=1}^{N} V_{ij}(t), \ t \geq T > 0.$$

Assume that

$$R_{\tilde{S}}^{out} = \bigcup_{i \in G_{\tilde{S}}^{out}} R_i^{out}$$

is the set of numbers of nodes – final receivers of flows, generated in the nodes, belonging to the set $G_{\tilde{S}}^{out}$. Divide the set $R_{\tilde{S}}^{out}$ onto two subsets, namely

$$R_{\tilde{S}}^{out} = R_{\tilde{S},int}^{out} \bigcup R_{\tilde{S},ext}^{out},$$

where $R_{\tilde{S},int}^{out}$ is the subset of nodes of $R_{\tilde{S}}^{out}$, belonging to $H_{\tilde{S}}$, and $R_{\tilde{S},ext}^{out}$ is the subset of nodes of $R_{\tilde{S}}^{out}$, belonging to addition to $H_{\tilde{S}}$ at the source network. The set $R_{\tilde{S},ext}^{out}$ will be called the domain of output influence of subsystem $\tilde{S}$ on network system [18]. External output strength of influence of nodes-generators of flows, belonging to the set $G_{\tilde{S}}^{out}$, on subnet $R_{\tilde{S},ext}^{out}$ determines by means of parameter

$$\alpha(t)_{\tilde{S}}^{out} = \sum_{i \in R_{\tilde{S},ext}^{out}} \xi_i^{out}(t) / N_{\tilde{S},ext}^{out},$$

where $N_{\tilde{S},ext}^{out}$ is equal to the number nodes-receivers of flows in addition to $H_{\tilde{S}}$ generated by the nodes of the subsystem $\tilde{S}$. Obviously, the value $\alpha_{\tilde{S}}^{out}(t)$ reflects the relative losses that await the system due to the blocking of nodes-generators of flows that are in an isolated zone of network for the period $[t-T, t]$, $t \geq T > 0$.

2. The total values of parameters of input influence of the nodes that are part of the isolated zone, come from the nodes that lie outside it, and allow us to evaluate the losses from the need to replace the nodes-receivers of flows that are in this zone. Denote by $R_{\tilde{S}}^{in}$ the set of all nodes-receivers of flows, belonging to the set $H_{\tilde{S}}$ and determine by means of parameter

$$\xi_{\tilde{S}}^{in}(t) = \sum_{i \in R_{\tilde{S}}^{in}} \xi_i^{in}(t) / s(\mathbf{V}(t))$$

the input influence strength of NS on subsystem $\tilde{S}$. Let us that

$$G_{\tilde{S}}^{in} = \bigcup_{i \in R_{\tilde{S}}^{in}} G_i^{in}$$

is a set of numbers of nodes-generator from which flows are sent to nodes belonging to the set $R_{\tilde{S}}^{in}$. Divide the set $G_{\tilde{S}}^{in}$ onto two subsets, namely

$$G_{\tilde{S}}^{in} = G_{\tilde{S},int}^{in} \cup G_{\tilde{S},ext}^{in},$$

where $G_{\tilde{S},int}^{in}$ is the subset of nodes of $G_{\tilde{S}}^{in}$, belonging to $H_{\tilde{S}}$, and $G_{\tilde{S},ext}^{in}$ is the subset of nodes of $G_{\tilde{S}}^{in}$, belonging to addition to $H_{\tilde{S}}$ at the source network. The set $G_{\tilde{S},ext}^{in}$ will be call the domain of input influence of NS on subsystem $\tilde{S}$ [18]. External input strength of influence of nodes-receivers of flows, belonging to the set $R_{\tilde{S}}^{in}$, on subnet $G_{\tilde{S},ext}^{in}$ determines by means of parameter

$$\alpha_{\tilde{S}}^{in}(t) = \sum_{i \in G_{\tilde{S},ext}^{out}} \xi_i^{in}(t) / M_{\tilde{S},ext}^{out}$$

where $M_{\tilde{S},ext}^{out}$ is equal to the number nodes-receivers of flows in $H_{\tilde{S}}$ generated by the nodes outside the subsystem $\tilde{S}$. Obviously, the value $\alpha_{\tilde{S}}^{in}(t)$ reflects the relative losses that await the system due to the blocking of nodes-receivers of flows that are in an isolated zone of network for the period $[t-T, t]$, $t \geq T > 0$.

3. The total values of flows betweenness parameters of elements that are part of isolated zone [18], provide the transit of flows through it and allow us to evaluate the losses from the need to redistribute flows by alternative paths that lie outside the isolated zone. Denote by $\xi_{\tilde{S}}^{tr}(t)$ the cost of all flows generated outside the subsystem $\tilde{S}$, which pass through it in transit. Then the value

$$\alpha_{\tilde{S}}^{tr}(t) = \xi_{\tilde{S}}^{tr}(t) / s(\mathbf{V}(t))$$

reflects the relative losses that await the system due to the need to redirect by alternative routes of transit flows that passed through isolated zone of network for the period $[t-T, t]$, $t \geq T > 0$.

Summarizing the values of parameters $\alpha_{\tilde{S}}^{out}(t)$, $\alpha_{\tilde{S}}^{in}(t)$ and $\alpha_{\tilde{S}}^{tr}(t)$, we can get a fairly adequate picture of losses that await the system, especially in the financial dimension, and make decisions that would minimize these losses. Of course, over time, the nodes-generator and nodes-receivers of flows that were in the isolated zone are replaced and alternative routes of motion of transit flows bypassing this zone are found. However, the appearance of such zones certainly destabilizes the system operation as a whole.

The railway transport system of the western region of Ukraine schematically reflected in fig. 1a (shows nodes with a structural degree 3 and more), part of which has become inaccessible (gray areas in Fig. 1b - 1d) [16]. From the given examples (fig. 1b, 1c) it follows that isolation of separate zones of CN can lead to loss of network connectivity. Sometimes such measures are forced and are an aim of isolation. However, in this case, the search for alternative routes of flow is significantly complicated or impossible. If the isolated zone does not lead to loss of CN connectivity (Fig. 1d, 1e), alternative routes are determined based on the following considerations.

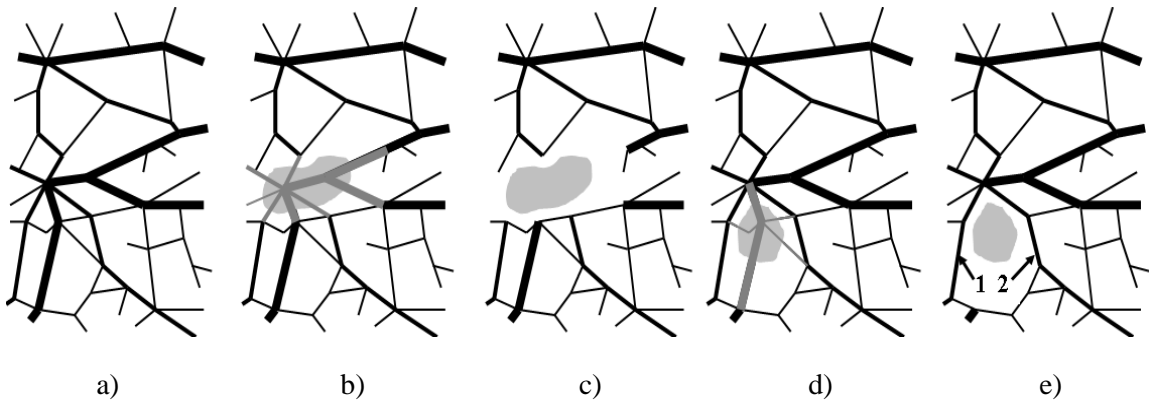

a)  b)  c)  d)  e)

Fig. 1. Isolated network zones and problem of finding alternative paths of flows motion

From a structural point of view, it is sufficient to find at least one shortest path around the isolated zone. From a functional point of view, we encounter the problem of redistribution of flows in all possible alternative ways. Obviously, the denser the CN, the greater the number of such paths. Thus, in the case of blocking node E (Fig. 2a), the search for alternative paths of flows motion bypassing this node causes significant problem. At the same time, having more connections in the network (Fig. 2b) removes these problems (paths A – D – B and A – C – B). That is, the most effective way to avoid the problems caused by the isolation of separate zones is the reserve of alternative paths of flows motion.

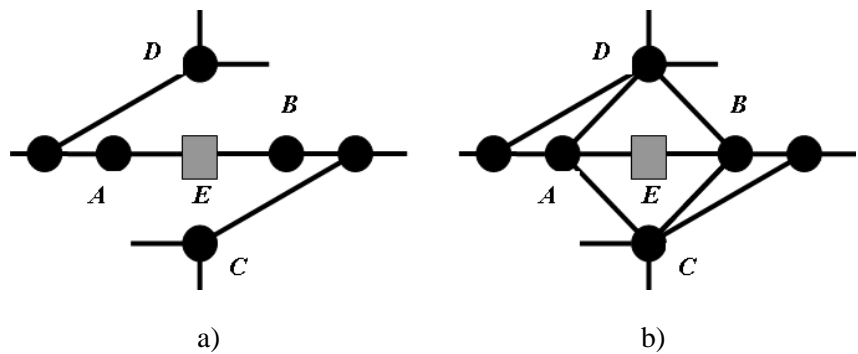

a)  b)

Fig. 2. Search for alternative paths depending on the network density

It is usually physically impossible to direct all transit flows on one shortest path due to the limited bandwidth of this path. In many real network systems, the separate flows have different priorities. This priority can be determined by the type of flow (international trains have a higher priority than suburban), its purpose, the importance of the generator or receiver of flow, and so on. That is, the possibility of redistribution of flows in alternative paths involves solving the following problems:

1) identification of possible alternative paths on the principle of "sending node – receiving node" and a list of routes taking into account their length - from the shortest to the longest allowable;

2) determination of the list of flows that need to be redirected, taking into account their priority;

3) determining the level of loading of each path and the volume of flows that can be directed by it;

4) the flow must not only go from a given initial to a given final node, but also pass as close as possible to the set of network nodes defined by its initial route.

Then the redistribution of flows is carried out according to the following algorithm:

1) the highest priority flows are sent by the shortest path until its capacity is exhausted (the threshold of its critical load is reached);

2) as the priority decreases, the flows are sent in other (longer) paths or with delay. That is, the higher the priority of flow, the shorter the path or the smaller the delay of its movement.

In some cases, there are problems that the some volume of flows has to be removed from the isolated zone (evacuation) or introduced into it in the shortest possible time (humanitarian aid, law enforcement units, etc.).

## 2. Granulation of complex network systems

The relevance of study of the problem of isolated network zones has increased especially with the deployment of the COVID-19 pandemic. Measures taken to combat the disease have effectively transformed the world into a network of isolated zones-countries, the movement of flows between which (especially human) due to the cessation or significant restriction of rail, air and road traffic has decreased by orders of magnitude. Moreover, as a result of introduction of such restrictions, many states have also become networks of isolated communities or separate settlements. The self-isolation of majority of citizens, caused by traffic restrictions, large fines for non-compliance with quarantine conditions, the closure of enterprises and work of people in remote access, has significantly reduced not only external but also internal flows in such isolated zones. With the constant network structure there was a kind of "granulation" of the system, which was divided into a hierarchy of successively isolated in terms of limiting the interaction of subsystems (Earth → countries → regions of countries → cities (neighborhoods of large cities) → citizens). These circumstances naturally led to a reduction in production and trade, the losses from which significantly accelerated and deepened the next financial and economic crisis. Thus, a new type of "granular" network systems has emerged, which has not been studied so far. Network granulation can be perceived as the transition of the system to another, critical mode of operation [19], and the process of its operation in this mode can be analyzed as a history of system behavior at certain intervals from pandemic beginning [20].

In the case of network granulation, the losses suffered by the respective system are caused not by blocking its individual component, but by restricting the movement of flows between all components as a whole. These system losses can be calculated sequentially by depicting the process of granulation and the corresponding reduction of flows volumes by the network using a temporal network, each layer of which $S^n = S(t_n)$, $n = 0,1,2,...$, sequentially reflects the next stage of granulation or a certain time stay of the NS at a given stage. If $s_0 = s(\mathbf{V}(t_0))$ reflects the total volume of network flows (in monetary terms) in the pregranular state $t_0$, then the ratio

$$P_n = s(\mathbf{V}(t_n))/s_0$$

gives a clear quantitative picture of the reduction of flows volumes caused by granulation of the network at moments of time $t_n$, $n=1,2,...$ The behavior of sequence $P_n$ allows us, in particular, to investigate the strengthening or weakening of the granulation mode of network.

## 3. Conclusions

Isolated zones often occur in complex network systems. Understanding the causes of nascency of such zones and their consequences allows us to minimize the losses that await the system, and faster return it to normal operation. The longer the granulation of network, the greater the damage to its structure and process of system functioning. The transition of NS to the "granular" state, which is characterized by partial isolation of all its components, creates a new kind of stability problems, the feature of which is the defeat not of its separate elements, but of the system as a whole. The development of scenarios of network system lesions that lead to its granulation, and means of timely counteraction to such lesions, is also an extremely important applied problem, for the theoretical and practical solution of which much effort will be spent.